\documentclass[10pt]{article}
\usepackage{graphicx}
\usepackage[margin=1.25in]{geometry}
\usepackage[usenames,dvipsnames]{color}
\usepackage{url}
\usepackage[colorlinks = true,
            linkcolor = blue,
            urlcolor  = blue,
            citecolor = blue,
            anchorcolor = blue]{hyperref}

 \usepackage[export]{adjustbox}

\newcommand\pubnumber{}
\newcommand\pubdate{\today}

\def\Title#1{\begin{center} {\LARGE #1 } \end{center}}
\def\Author#1{\begin{center}{ \sc #1} \end{center}}
\def\Address#1{\begin{center}{ \it #1} \end{center}}

\newcommand\pubblock{\rightline{\begin{tabular}{l} \pubnumber\\
         \pubdate \end{tabular}}}
\newenvironment{Abstract}{\begin{quotation} \begin{center}
                       ABSTRACT
     \end{center}\bigskip  }{\end{quotation}}

\def\ie{{\it i.e.}}
\def\eg{{\it e.g.}}

\def\beq{\begin{equation}}
\def\eeq#1{\label{#1}\end{equation}}
\def\eeqn{\end{equation}}

\newenvironment{Eqnarray}%
   {\arraycolsep 0.14em\begin{eqnarray}}{\end{eqnarray}}
\def\beqa{\begin{Eqnarray}}
\def\eeqa#1{\label{#1}\end{Eqnarray}}
\def\eeqan{\end{Eqnarray}}

\let\bar=\overbar

\def\lsim{\mathrel{\raise.3ex\hbox{$<$\kern-.75em\lower1ex\hbox{$\sim$}}}}
\def\gsim{\mathrel{\raise.3ex\hbox{$>$\kern-.75em\lower1ex\hbox{$\sim$}}}}

\def\del{\partial}
\def\Dslash{\not{\hbox{\kern-4pt $D$}}}
\def\dslash{\not{\hbox{\kern-2pt $\del$}}}
\def\pslash{\not{\hbox{\kern-2pt $p$}}}
\def\ETmiss{\not{\hbox{\kern-4pt $E$}}_T}

\def\Dlr{\mathrel{\raise1.5ex\hbox{$\leftrightarrow$\kern-1em\lower1.5ex\hbox{$D$}}}}

\def\Pl{{\mbox{\scriptsize Pl}}}

\def\GUT{{\mbox{\scriptsize GUT}}}

\def\MSB{{\bar{M \kern -2pt S}}}
\def\msb{{\bar{\scriptsize M \kern -1pt S}}}

\def\drb{{\bar{\scriptsize D \kern -1pt R}}}

\def\GeV{{\rm GeV}}

\newcommand\snowmass{\begin{center}\rule[-0.2in]{\hsize}{0.01in}\\\rule{\hsize}{0.01in}\\
\vskip 0.1in Submitted to the  Proceedings of the US Community Study\\ 
on the Future of Particle Physics (Snowmass 2021)\\ 
\rule{\hsize}{0.01in}\\\rule[+0.2in]{\hsize}{0.01in} \end{center}}

\newcommand\zosoir{$\includegraphics[width=0.09\hsize,valign=c]{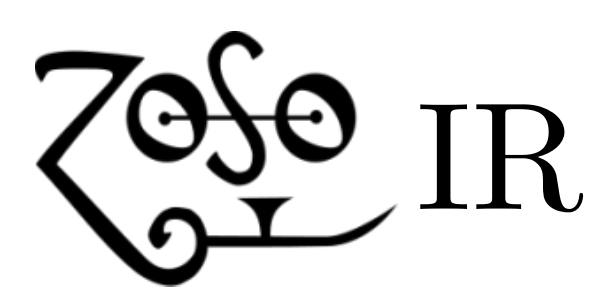}$}
\newcommand\zosouv{$\includegraphics[width=0.09\hsize,valign=c]{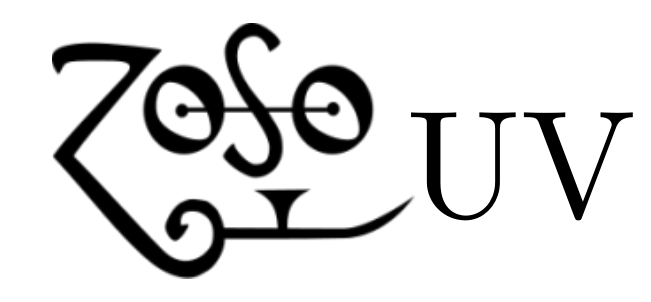}$}

\begin{document}

\pubblock

\Title{Effective field theories of gravity and compact binary dynamics:  A Snowmass 2021 whitepaper}
\bigskip 

\Author{Walter D. Goldberger}

\medskip

\Address{ Physics Department, Yale University, New Haven, CT 06520, USA}

\medskip

 \begin{Abstract}

\noindent In this whitepaper, I describe modern applications of effective field theory (EFT) techniques to classical and quantum gravity, with relevance to problems in astrophysics and cosmology.     As in applications of EFT to high-energy, nuclear, or condensed matter physics, the Wilsonian paradigm based on decoupling of short distance scales via renormalization group evolution remains a powerful organizing principle in the context of gravity.  However, the presence of spacetime geometry brings in new elements (\eg~ non-trivial time-dependence, cosmological or black hole event horizons) which necessitate the introduction of novel field theoretic methods not usually encountered in applications of EFTs to physics at the energy and intensity frontiers.  After a brief overview of recent developments in the application of EFT methods to gravity, I will focus on the EFT description of compact binary dynamics, including an overview of some of its applications to the experimental program in gravitational wave detection at LIGO/VIRGO and other observatories.

\end{Abstract}

\snowmass

\section{Modern EFTs of gravity}

Even though quantum gravity remains mysterious in the ultraviolet (UV), \ie~at energy scales near the Planck mass 
$$m_{\Pl}=1/\sqrt{32\pi G_N}\sim 10^{19} \GeV,$$
 its effects at long distance scales are much better understood.    It has been well appreciated~\cite{Weinberg:1964ew,Weinberg:1965rz}  since at least the 1960s that, regardless of the microscopic structure of quantum gravity, in the infrared (IR) any quantum theory whose spectrum contains massless helicity-two particles (gravitons) coupled to matter must be described , on the basis of general principles\footnote{\eg~ Poincare invariance, unitarity and cluster decomposition of the $S$-matrix, etc.} by a quantum field theory whose Lagrangian coincides with that of general relativity
\begin{equation}
\label{eq:eh}
{\cal L} = - 2 m_{\Pl}^2\sqrt{g} R(x)+ {\cal L}_{matt}+\cdots
\end{equation}
at sufficiently low energies, up to higher-order curvature corrections which are suppressed at energies below the Planck scale.

Expanding out Eq.~(\ref{eq:eh}) around a flat spacetime background
$$
{g}_{\mu\nu} = \eta_{\mu\nu} + {h_{\mu\nu}\over m_{Pl}},
$$
generates an infinite tower of graviton self-interactions of the schematic form
$$
{\cal L} = ({\partial h})^2 + {1\over m_{Pl}} h ({\partial h})^2+ {1\over m^2_{Pl}} h^2 ({\partial h})^2+\cdots.
$$
whosequantization was achieved already in~\cite{Gupta:1952zz,Feynman:1963ax,DeWitt:1967yk,DeWitt:1967ub,DeWitt:1967uc}.   In particular, DeWitt~\cite{DeWitt:1967uc} derived the Feynman rules in Lorentz covariant form, and used them to obtain predictions for the tree-level (\ie~${\cal O}(G_N))$ $S$-matrix in two-body collisions involving asymptotic states with at least one graviton.   The same formalism is also capable of incorporating any number of arbitrarily soft gravitons in the initial or final state.  As Weinberg showed in a classic paper~\cite{Weinberg:1965nx}, the IR divergences due to the emission of many soft gravitons exponentiate and factorize from the $S$-matrix, and are resolved in the same manner as in the case of soft photon emission in QED.    One-loop quantum corrections were first calculated out in~\cite{tv}, which found that in the absence of matter, quantum gravity is one-loop UV finite, with the first non-trivial UV divergence appearing at two-loop order~\cite{Goroff:1985th}, requiring the addition of local counterterms which are cubic in the Riemann curvature.

By power counting, one expects UV divergences to arise at all loop orders, requiring the addition of an infinite number of higher curvature diffeomorphism-invariant counterterms to Eq.~(\ref{eq:eh}), with arbitrary coefficients scaling as more powers of $1/m_{Pl}$.   From a  modern point of view, the non-renormalizability of the Lagrangian in Eq.~(\ref{eq:eh}) is understood to mean that this theory is an \emph{effective field theory}, valid only at sufficiently long distances:    While the UV divergences can be treated systematically order by order in $(E/m_{Pl})^2\ll 1$, the quantum field theory defined by Eq.~(\ref{eq:eh}) plus higher curvature terms should be regarded as having limited predictive power.   Effects that depend non-analytically on kinematic invariants are calculable in the EFT, but local (analytic) effects are encoded in unknown Wilson coefficients that can be determined through matching to a more complete UV theory of quantum gravity or, in principle, extracted from experiments.

An example due to Donoghue~\cite{Donoghue:1993eb}, who was first to explicitly emphasize the EFT interpretation of canonically quantized general relativity~\cite{Donoghue:1993eb,Donoghue:1994dn},  are the quantum corrections to the gravitational potential between test particles\footnote{This result~\cite{Bjerrum-Bohr:2002gqz} incorporates graviton loops, and not loops of other massless Standard Model particles.    See also~\cite{Muzinich:1995uj} for complementary approaches based on other definitions of the non-relativistic gravitational potential.},
\begin{equation}
\label{eq:vhbar}
V_{{\cal O}(\hbar^1)}({\bf x}) = c(\mu)\delta^3({\bf x}) -{41\over 10\pi} {G_N\hbar\over r^2}\cdot \left({G_N m_1 m_2\over r}\right).
\end{equation}
The $O(1/r^3)$ correction to the potential is determined by the Fourier transform of a term $\sim\log{\vec q}^2/\mu$ in the scattering amplitude of non-relativistic particles, and therefore gives rise to a well-defined long distance prediction, while the renormalization scale dependence is absorbed into a counterterm $c(\mu)$ that encodes sensitivity to (unknown) short distance physics, which manifests itself as a short-range (contact) potential between the point masses.   For reviews of the EFT approach to quantum gravity, see~\cite{Donoghue:1995cz,Burgess:2003jk}.

It is therefore fair to say that, at least in the context of flat spacetime, the IR behavior of quantum gravity is well under theoretical control.   Of course, the predictions of this theory will not be tested experimentally any time soon~\cite{Adelberger:2022sve},  \eg~ for test masses separated by a distance of $r\sim1\mbox{ mm}$, the corrections to the classical potential from Eq.~(\ref{eq:vhbar}) are down by a factor of $10^{-64}$ relative to the Newtonian gravitational potential.    However, even if the long distance experimental consequences of quantum gravity are too minuscule to be relevant, the EFT description of gravity in principle yields sharp predictions for physical observables\footnote{Despite the smallness of $E^2/m^2_{Pl}$ corrections to macroscopic observables, it has been theoretically very fruitful in recent years to push the state of the art in computing the $S$-matrix of low energy quantum gravity via field theoretic techniques.    Much progress has been made in the last decade toward understanding the structure of such observables, leading to new connections to perturbative gauge theory and to the development of efficient techniques for calculating observables of phenomenological relevance.   This subject is reviewed in refs.~\cite{Bern:2019prr,Adamo:2022dcm}.}.

The perturbative quantization of Eq.~(\ref{eq:eh}) around non-trivial background spacetimes $\langle g_{\mu\nu}\rangle\neq\eta_{\mu\nu}$ is also well established.    As long as the curvature of this background is sufficiently small relative to $m_{Pl}^2$, effective field theory should still yield reliable results at sufficiently long distances.   For instance Hawking's~\cite{Hawking:1974sw} analysis of free quantum fields propagating near the horizon of a black hole is justified only if effective field theory is valid\footnote{The question of wether effective field theory breaks down~\cite{Hawking:1976ra} in the IR, at late time scales of order the black hole's lifetime $t_{Page} \sim M_{BH}^3/m_{Pl}$~\cite{Page:1976df}  is not fully settled, although recent developments~\cite{Almheiri:2020cfm} suggest that non-perturbative effects in EFT can reproduce Page curve.}, which requires $M_{BH}\gg m_{Pl}$.   Similarly, the reason that it is possible to theoretically predict the effects of early universe physics (\eg~ a nearly deSitter period of inflation at Hubble scales $H\lsim M_{\GUT}\sim 10^{17}\mbox{GeV}$) on CMB correlations is that there is a hierarchy $H/m_{Pl}\ll 1$ that allows for the decoupling of unknown quantum gravity effects.  If inflation happens at a high energy scale, the effects of quantum gravity on CMB correlations need not be hopelessly small.    In this case, the quantization of the gravitational field as predicted by EFT has experimental consequences:    measurements of a primordial $B$-mode pattern of polarizations in the CMB can become sensitive~\cite{Seljak:1996gy,Kamionkowski:1996zd} to the graviton propagator $\langle h_{ij} h_{kl}\rangle\sim H^2/m_{Pl}^2$ during inflation~\cite{Abbott:1984fp}.   Even three-point tensor non-Gaussianities~\cite{Maldacena:2002vr,Maldacena:2011nz} $\langle h h h \rangle$, which are directly related to the self-couplings of the graviton are potentially measurable, opening the way to a host of potential new signatures~\cite{Arkani-Hamed:2015bza} of UV physics in the CMB.

The modern applications of EFTs to gravitational systems in astrophysics and cosmology usually involve the (spontaneous) breaking of Poincare invariance, either due to the presence of non-trivial background spacetime curvature or of dynamical sources such as black holes or other extended objects of diverse dimensionality (strings, branes, etc).  In these settings, there is usually a hierarchy of momentum or energy scales 
$$
m_{\Pl}\gg \Lambda_{UV}\gg \Lambda_{IR},
$$
between the scale $\Lambda_{UV}$ set by the sources that couple to gravitons and the typical low energy scale $\Lambda_{IR}$ set by the kinematics of the observables of interest.   As in conventional applications outside of gravity, it is convenient to disentangle the hierarchy of scales by constructing a tower of gravitational theories~\cite{Goldberger:2006bd} containing only the dynamical degrees of freedom relevant at each scale.    The advantages for doing so are the same as in conventional EFTs:   
\begin{itemize}
\item \emph{Power counting}:   The Wilson coefficients of local operators in the effective Lagrangian depend only on $\Lambda_{UV},$ so that power counting in the expansion parameter $\Lambda_{IR}/\Lambda_{UV}\ll 1$ is manifest.
\item \emph{Analyticity of short distance contributions}:   UV effects are in one-to-one correspondence with local operators in the effective Lagrangian.   Thus at any given order in $\Lambda_{IR}/\Lambda_{UV},$ the most general Lagrangian that is consistent with the symmetries of the relevant degrees of freedom at $\Lambda_{IR}$ necessarily describes the UV physics in a model-independent way.   For suitably defined observables, these short distance contributions depend analytically on the kinematics.
\item  \emph{Renormalization group (RG) evolution}:   Non-analytic contributions, in the form of large logarithms $\ln \Lambda_UV/\Lambda_{IR}\gg 1$, can be understood as the RG evolution of the EFT Wilson coefficients from a matching scale $\mu\sim \Lambda_{UV}$ where the EFT is no longer a complete description of the physics down to the IR at a scale $\mu\sim\Lambda_{IR}$.  The scaling dimensions of the Wilson coefficients are calculable in the EFT and non-analytic effects in $\Lambda_{IR}/\Lambda_{UV}$ are therefore universal.
\end{itemize}

On the other hand, in the presence of gravity there can be conceptual twists not usually encountered in applications of EFTs to scattering in high energy or nuclear physics.   One clear difference is the nature of what are the well-defined (i.e. diffeomorphism invariant) observables of the theory.  This depends both on the precise nature of the asymptotic  boundary of spacetime at infinity, as well as on the choice of boundary conditions which define the asymptotic states.  In turn, these properties depend on the vacuum, which may not be fully Poincare invariance due to the presence of spacetime curvature or of dynamical sources.   The possibility that the background possesses either cosmological or black hole event horizons also introduces a set of issues not normally encountered in applications of EFTs:   stimulated emission of particles and amplification of quantum fluctuations, dissipation, and UV/IR mixing, in the sense of stretching of short distance modes (as in inflationary cosmology) or blueshifting of soft quanta by the horizon of a black hole.

The question of how to modify the rules of EFT in  such spacetime backgrounds has been an active area of research in recent years, motivated by problems in cosmology and astrophysics \footnote{Similar questions also show up in applications of EFT to condensed matter, for instance how scale separation works in non-equilibrium systems.    See the Snowmass whitepaper~\cite{Brauner:2022rvf} for a review and a complete set of references.}.   Novel ideas and techniques have emerged , resulting in a variety of new ``designer'' EFTs to describe, \eg~ single field~\cite{Cheung:2007st,Weinberg:2008hq} and multi-field~\cite{Senatore:2010wk} early universe inflation, large scale structure~\cite{Baumann:2010tm,Carrasco:2012cv,Porto:2013qua} in the matter dominated era, and dark energy~\cite{Gubitosi:2012hu}.     The question of UV dependence of predictions of inflation for CMB correlations has been addressed using EFT reasoning in refs.~\cite{Kaloper:2002uj,Burgess:2002ub,Burgess:2003zw,Burgess:2009ea,Burgess:2010zq}, while the sensitivity to general initial conditions was analyzed in an EFT with Schwinger-Keldysh~\cite{KS} boundary conditions in refs.~\cite{Collins:2005nu,Agarwal:2012mq} (the necessity for in-in closed time path contours in the evaluation of cosmological correlators was emphasized in~\cite{Weinberg:2005vy}).   Loop corrections to cosmological correlators and the issue of IR divergences (secular growth in time) has been addressed using EFT methods (power counting, RG flows, etc.) in~\cite{Weinberg:2005vy,Weinberg:2006ac,Burgess:2009bs,Senatore:2009cf,Giddings:2010nc,Marolf:2010zp,Burgess:2014eoa,Burgess:2015ajz,Gorbenko:2019rza,Cohen:2020php,Green:2020txs}.   These references only scratch the surface, see the Snowmass whitepaper~\cite{Flauger:2022hie}, for a review of the role of fundamental theory in cosmology, and~\cite{Cabass:2022avo} more specifically for a summary of effective field theories.

 A modern theme~\cite{Adams:2006sv,Camanho:2014apa,Arkani-Hamed:2020blm} in theoretical physics is the idea of ``bootstrapping'' low energy dynamics using a small set of general principles that originate from consistency in the UV.   This approach is in some ways complementary to EFT reasoning, and has been fruitfully applied to field theories containing gravity.   See the Snowmass whitepaper~\cite{Baumann:2022jpr} for a review of the ``cosmological bootstrap'' approach to CMB correlators,  and~\cite{deRham:2022hpx,Draper:2022pvk} for reviews of the more general program of constraining low energy EFTs of gravity using UV consistency conditions.   The study of EFTs of gravity in Anti-deSitter backgrounds~\cite{Heemskerk:2009pn}  has also seen huge progress thanks to its connection, via the AdS/CFT correspondence~\cite{Maldacena:1997re} to conformal field theory in Minkowski spacetime.   It is therefore possible to extract information about long distance quantum gravity in AdS from bootstrap constraints on CFTs.   See the Snowmass whitepapers~\cite{boots} for a review and further references.

 EFTs of gravity have also played a role in other areas of astrophysics.   The EFT approach to the dynamics of gravitationally bound compact objects, with applications to gravitational wave emission, was initiated in ref.~\cite{Goldberger:2004jt}.   It exhibits many of the elements that are common in the modern applications of EFTs to gravitational physics, from Schwinger-Keldysh~\cite{KS} boundary conditions in the path integral, to spacetime-dependent Wilson coefficients and UV divergences, that result in novel types of RG flows, to new forms of IR behavior, including absorption or amplification of long distance degrees of freedom by horizons, etc.   

 This whitepaper is intended as a brief review of EFT methodology in gravitational physics that has been the subject of theoretical research in recent the last decade or so.   The discussion will be framed in the context of the ``Wilsonian'' approach to compact binary dynamics introduced in~\cite{Goldberger:2004jt}, as this field theory of gravity serves to illustrate the new theoretical structures that arise generally when applying EFT ideas to systems that include gravity.   Detailed reviews of EFT for other systems that include gravity can be found in the references, \eg~~\cite{Flauger:2022hie,Cabass:2022avo}.

\section{EFT of gravity for compact binary dynamics}

\begin{figure}
\begin{center}
\includegraphics[width=0.60\hsize]{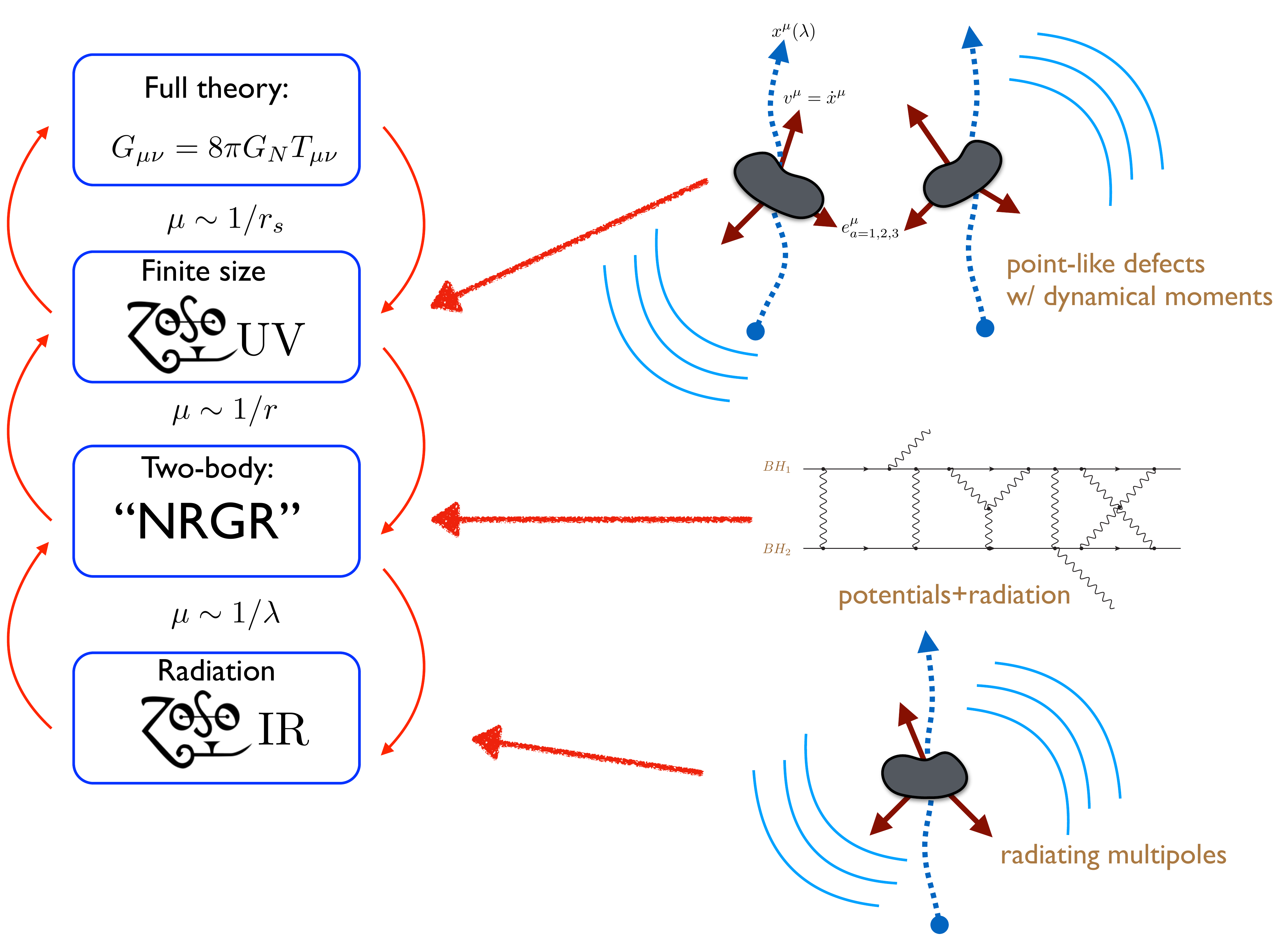}
\end{center}
\caption{Tower of gravity EFTs for non-relativistic compact bound states.}
\label{fig:tower}
\end{figure}

Understanding the dynamics of gravitationally bound compact (black hole or neutron star) binaries is crucial to the experimental program in gravitational wave detection~\cite{GW}.   As the compact objects orbit one another, they release energy in the form of gravitational radiation, eventually merging and coalescing into a (presumably) final stationary black hole state.   In the intermediate stages, when the objects are separated by distances of order the Schwarzschild radius $r_s=2 G_N M$, gravity is in its strongly coupled, non-linear phase, and the binary problem requires the toolset of numerical general relativity~\cite{Lehner:2014asa}.    On the other hand, at sufficiently early times the dynamics can be treated analytically, as a perturbative expansion in a small parameter
$$v^2 \sim {r_s\over r} \ll 1,$$
set by the orbital parameters of the system.   The post-Newtonian (PN) limit refers to the systematic solution of the Einstein equations of the compact binary order-by-order as an expansion in the relative velocity in the kinematic regime $v\ll 1$.  See~\cite{blanchet,Schafer:2018kuf} for reviews of classical approaches to the PN problem as well as more complete references.

The PN limit of binary dynamics has a natural formulation in the language of EFTs~\cite{Goldberger:2004jt} (for reviews, see~\cite{Goldberger:2006bd,Foffa:2013qca,Rothstein:2014sra,Porto:2016pyg,Levi:2018nxp}.  The system exhibits a hierarchy of widely separated length scales
$$
r_s\lsim {\cal R} \ll r \sim v^{-2} r_s \ll \lambda \sim v^{-1} r
$$
ranging from the gravitational radius of the system $r_s\sim 2 G_N M,$ to the physical radius ${\cal R}$ of the compact objects (for black holes we define ${\cal R}=r_s$ while for other compact objects ${\cal R}/r_s\gsim{\cal O}(1)$), to the orbital radius $r\sim r_s/v^2$, to the typical wavelength $\lambda\sim r/v$ of the gravitational radiation emitted by the binary.   Furthermore, these scales are \emph{correlated}, in the sense that scale ratios are given by powers of the expansion parameter $v$.   Therefore at any given order in the PN expansion\footnote{The convention is to call a term ``$n$PN'' if it contributes at order $v^{2n}$, where $n\geq 0$ is integer or half integer.}, qualitatively distinct effects, due to physics at different length scales, have to be taken into account.

To disentangle the corrections coming from physics at all these different scales, one can construct a tower of EFTs of gravity, as sketched in Fig~\ref{fig:tower}.   For our purposes the microscopic ``full theory" in the UV  is given by classical general relativity, coupled to some source $T^{\mu\nu}$ of conserved energy-momentum that describes the internal structure of each compact object.    For example, if we assume that a neutron star is a configuration of nuclear matter which is in a hydrodynamical phase, the source $T^{\mu\nu}$ would take the form of a perfect fluid with thermodynamic equation of state taken as an input from nuclear physics/QCD.    Additional fluid transport coefficients (\eg~ bulk and shear viscosity) can also be included in $T^{\mu\nu}$ if necessary and are suppressed at long distances .   For the sake of brevity, we will restrict ourselves in this paper to the case of black hole binaries, so that the full theory is simply the classical vacuum Einstein equation $R_{\mu\nu}=0$ in the sector containing two black holes (the spacetime has a topologically disconnected event horizon in the far past).

In the EFT, the compact objects themselves are treated as dynamical point-like defects (worldlines) that carry internal degrees of freedom coupled to gravity.   This EFT efficiently describes finite size effects originating from the internal structure of the orbiting compact objects, \eg~ tidal deformations induced by external gravitational fields, or dissipation of energy across the BH horizon.    A review of the construction of this finite size EFT  will be postponed until sec.~\ref{sec:UV}, focusing first on the EFT description of non-relativistic two-body dynamics at the orbital scale $r$ and radiation at scales $\sim r/v$ in sec.~\ref{sec:NRGR}.

In particular, if we temporarily assume that the internal dynamics is gapped at a distance scale ${\cal R}\ll r$,  the only light degrees of freedom are the Goldstone modes associated with the spontaneous breaking of local Poincare symmetry by the presence of the compact object.   Thus, up to gauge redundancy, in the point particle limit each binary constituent is described by a worldline $x^\mu(\tau)$ and a spin degree of freedom $S^{\mu\nu}(\tau)=-S^{\nu\mu}(\tau)$ localized on each worldline.   The dynamics then follows from a worldline action that couples $x^\mu$ and $S^{\mu\nu}$ to the gravitational field $g_{\mu\nu}$, whose general form is determined by general principles:  (1) invariance under diffeomorphisms of $g_{\mu\nu}$ or reparameterizations of the worldline, and (2) smooth  $g_{\mu\nu}\rightarrow \eta_{\mu\nu}$ limit.  

These principles result in a Lagrangian that is generically a sum over an infinite number of monomial invariants of $g_{\mu\nu}$, $dx^\mu/d\tau,$ $S^{\mu\nu}$ and their derivatives.   Each  term carries a Wilson coefficients that, by dimensional analysis, scales as a non-negative integer power of the scale ${\cal R}$.   By writing down the most general theory consistent with the above principles, with arbitrary\footnote{Note that the Wilson coefficients cannot be completely arbitrary.    In EFTs that emerge as low energy limits of UV complete theories, quantum mechanical unitarity and causality imply non-trivial constraints on parameter space~\cite{Adams:2006sv,Camanho:2014apa}.  Understanding the constraints on EFTs that couple to gravity has been the active focus of recent research, see~\cite{deRham:2022hpx} for a detailed guide to the literature.} Wilson coefficients, we are necessarily parameterizing the most general compact object that can exist in the full UV theory.  In the regime of validity of the EFT, a given Wilson coefficient yields a contribution to an observable that scales as a definite power of ${\cal R}/r\ll 1$, so that in practice we only need to know a finite number of parameters in order to make a prediction with finite precision.

Organizing the EFT power counting as an expansion in derivatives of $g_{\mu\nu}$, the most general worldline theory takes the form
 \begin{equation}
 \label{eq:pp}
S_{pp} = -m\int d\tau + c_E \int d\tau E_{\mu\nu} E^{\mu\nu} + c_B\int d\tau B_{\mu\nu} B^{\mu\nu} +\cdots.
\end{equation}
By the Equivalence Principle, at leading (zeroth) order in derivatives, compact object dynamics is universal, with worldlines that follow timelike geodesics of $g_{\mu\nu}$.   The first deviations from test particle motion arise at second order in derivatives of the metric, and involve the square of the ``electric", $E_{\mu\nu}$, and ``magnetic" $B_{\mu\nu}$ components of the Weyl tensor\footnote{Terms involving the Ricci curvature can be removed by field redefinitions of the metric.   We are assuming here parity invariance which forbids a worldline term of the form $\int d\tau E_{\mu\nu} B^{\mu\nu}$, whose phenomenology has been recently studied in ref.~\cite{Modrekiladze:2022ioh}.}.    The coefficients $c_{E,B}$ measure the leading (quadrupolar) tidal response of the compact object to an external gravitational field, and by dimensional analysis are expected to scale as $c_{E,B}\sim {\cal R}^5/G_N$.    As such, $c_{E,B}$ provide a gauge invariant definition for the $\ell=2$ static tidal ``Love numbers" which characterize the gravitational response~\cite{Goldberger:2006bd,Goldberger:2007hy,Goldberger:2005cd}.   These Love numbers are in general dependent on the form of the equation of state of the compact star, and have been computed  first in refs.~\cite{Flanagan:2007ix,Hinderer:2007mb,Hinderer:2009ca} in successively increasing levels of physical detail.   Because of the scaling with radius, the effects of $c_{E,B}$ on binary dynamics are expected to scale as $({\cal R}/r_s)^5\times v^{10}$, so that although formally a 5PN correction, finite size effects are enhanced for objects that are less compact than black holes\footnote{For the case of Schwarzschild black holes, calculations in full general relativity~\cite{Binnington:2009bb,Damour:2009va}, when matched to the point particle EFT~\cite{Kol:2011vg}, indicate that $c^{BH}_{E,B}=0$.  The vanishing of the static linear response of black holes has been extended to higher multipoles~\cite{Pani:2015hfa} and to non-zero spin in~\cite{Pani:2015hfa,LeTiec:2020spy,Chia:2020yla,LeTiec:2020bos}.   See~\cite{Barack:2018yly} for further references.    In the absence of some hidden symmetry, the vanishing of the Wilson coefficients of the point particle EFT of black holes is in tension with expectations based on naturalness criteria as well as sum rules that follow from causality~\cite{GnR,Rothstein:2014sra,Porto:2016zng}.  It remains an open question wether there is a fundamental (\eg~ based on some underlying symmetry) principle that explains the vanishing of black hole static Love numbers, see~\cite{Charalambous:2021mea,Charalambous:2021kcz,Hui:2021vcv,Hui:2022vbh} for some recent proposals.   Note though that the \emph{non-static} AC response ($\omega\neq 0$) of black holes is non-zero.  See sec.~\ref{sec:UV} for further discussion.} ${\cal R}\gsim r_s$.  This observation~\cite{Flanagan:2007ix} provides strong motivation for carrying out analytical PN calculations to at least $O({\cal} v^{10})$ where such tidal  effects start to appear.

In Eq.~(\ref{eq:pp}), I have omitted the infinite tower of higher order curvature invariants that characterize additional finite size effects, suppressed by higher powers of ${\cal R}/r\ll 1$.   Also not displayed in Eq.~(\ref{eq:pp}) are the terms in the action necessary to keep track of the evolution of the compact object's spin $S^{\mu\nu}$.    The inclusion of spin into the EFT framework, which is beyond the scope of this review, is of crucial importance for phenomenology,   This was pioneered by Porto in ref.~\cite{Porto:2005ac}, generalizing the phase space formulation of Regge and Hanson~\cite{Hanson:1974qy} to curved spacetime.   See~\cite{Delacretaz:2014oxa} for a coset formulation of the general relativistic spinning particle emphasizing the role of non-linearly realized local Poincare invariance.     

For gravitational wave detection, the relevant observable is the waveform $h_{\mu\nu}=g_{\mu\nu}-\eta_{\mu\nu}$ measured by observers at future null infinity ($r\rightarrow\infty$ and fixed retarded time).   In the EFT of gravitons coupled to compact objects described by Eq.~(\ref{eq:pp}), this corresponds to an expectation value
$$
\langle in| h_{\mu\nu}(x)|in\rangle,
$$
evaluated in the initial state of the radiation field and of the binary constituents.   As was first emphasized by Galley and Tiglio~\cite{Galley:2009px}, because we are holding the initial state fixed but summing over all final states of the coalescing binary, the appropriate formalism for setting up perturbation theory is the Schwinger-Keldysh~\cite{KS}  closed time path (CTP) or ``in-in" functional integral.   This is analogous to the situation in cosmology, where late time correlations are measured in a given initial state~\cite{Weinberg:2005vy}.    In order to calculate $\langle in| h_{\mu\nu}(x)|in\rangle$, a convenient generating function for observables is the in-in effective action induced by integrating out gravitons in the presence of fixed worldlines $x_{a=1,2}^\mu(\tau)$ (and their spins), as well as a background gravitational field ${\bar g}_{\mu\nu} = \eta_{\mu\nu} + {\bar h}_{\mu\nu}$,
\begin{equation}
\label{eq:inin}
e^{i\Gamma[x_a,{\bar g},{\tilde x}_a {\tilde{\bar g}}]} =\int {\cal D} h_{\mu\nu}(x) {\cal D}{\tilde h}_{\mu\nu}(x) e^{i S[{\bar g},h,x_a] - i S[{\tilde {\bar g}},{\tilde h},{\tilde x}_a]}.
\end{equation}
We denote the classical action by $S[{\bar g},h,x_a]=S_{EH}[{\bar g}+h] + S_{pp}[{\bar g}+h,x_a]+S_{GF}[{\bar g},h]$, which includes a suitable gauge-fixing term, in practice chosen to preserve gauge invariance with respect to diffeomorphisms acting on the background field ${\bar g}_{\mu\nu}$.

\begin{figure}
\begin{center}
\includegraphics[width=0.50\hsize]{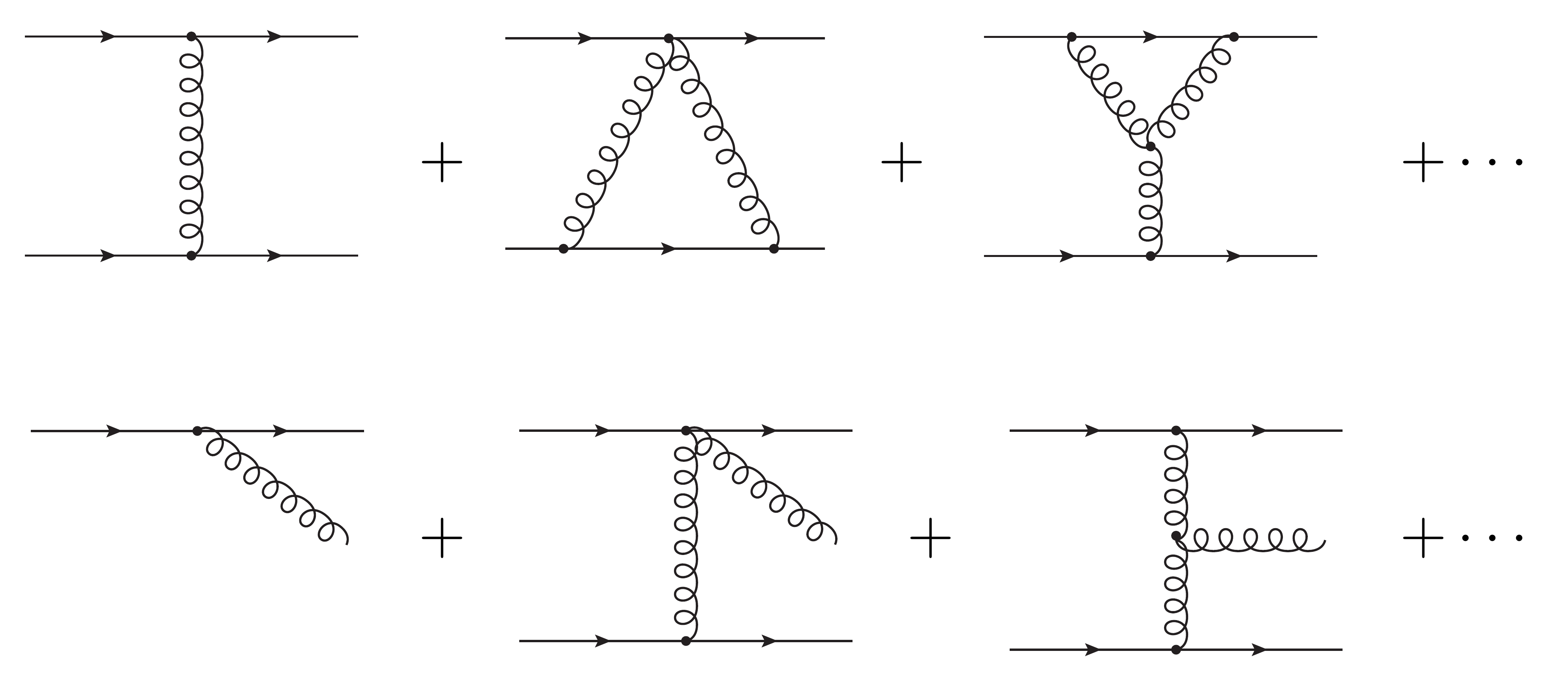}
\end{center}
\caption{Feynman diagram expansion of the in-in action.   External lines correspond to insertions of the background field ${\bar h}_{\mu\nu}$,  ${\tilde {\bar h}}_{\mu\nu}$.   Each diagram shown stands for a sum of contributions from insertions on either side of the Schwinger-Keldysh closed time contour.}
\label{fig:LOdiags}
\end{figure}

From the in-in action $\Gamma\left[x_a,{\bar g},{\tilde x}_a,{\tilde{\bar g}}\right]$, we obtain the classical equations of motion for the gravitationally interacting worldlines by extremization,
$$
\left.{\delta\over\delta x^\mu_a}\Gamma[x_a,{\bar g},{\tilde x}_a {\tilde{\bar g}]}\right|_{x_a={\tilde x}_a;{\bar g}={\tilde{\bar g}}=\eta}\equiv 0.
$$
The solution to these equations of motion is then inserted into the energy-momentum pseudotensor $\tau_{\mu\nu}(x)$, defined as the variation with respect to the background field
$$
\tau_{\mu\nu} = {2\over \sqrt{{\bar g}}} {\delta\over \delta {\bar g}^{\mu\nu}} \left.\Gamma[x_a, {\bar g}, {\tilde x}_a \tilde{\bar g}]\right|_{x_a={\tilde x}_a;{\bar g}={\tilde{\bar g}}=\eta}.
$$
By the diff invariance of the background field, this pseudo-tensor is conserved on-shell, $\partial_\nu \tau^{\mu\nu}=0$, but dependent on the gauge fixing term $S_{GF}[{\bar g},h]$.   This pseudo-tensor has a direct relation to the quantum mechanical amplitude for the binary system to emit an on-shell graviton of definite momentum $k^\mu$ ($k^2=0$),
$$
{\cal A}(k) = \epsilon_{\mu\nu}(k) {\cal A}^{\mu\nu}(k) =  -{1\over 2 m_{Pl}} \int d^4 x e^{ik\cdot x} \epsilon_{\mu\nu}(k) \tau^{\mu\nu}(x).
$$
In turn, this on-shell amplitude has a simple relation to the waveform, once a gauge for the background field has been chosen.   \eg~, in deDonder gauge, the waveform at  future null infinity is
$$
\lim_{r\rightarrow \infty} \langle in| h_{\mu\nu}(x)|in\rangle = {4 G_N\over r} \int {d\omega\over 2\pi} e^{-i\omega t} \left[{\cal A}^{\mu\nu}(k)-{1\over 2} \eta^{\mu\nu} {\cal A}^\rho{}_\rho(k)\right],
$$
where the on-shell momentum is $k^\mu=\omega(1,{\vec x}/r)$.    

The effective action $\Gamma\left[x_a,{\bar g},{\tilde x}_a,{\tilde{\bar g}}\right]$ admits a formal perturbative expansion in powers of $G_N$.  The terms in the perturbative expansion can be organized by drawing  in-in Feynman diagrams with internal $h_{\mu\nu},{\tilde h}_{\mu\nu}$ graviton lines coupled to classical worldline sources as well as to external background gravitons ${\bar h}_{\mu\nu},{\tilde {\bar h}}_{\mu\nu}$, see Fig.~\ref{fig:LOdiags}.   At the classical level, only diagrams with at most one external ${\bar h},{\tilde {\bar h}}$ are relevant.

In order to solve the \emph{fully relativistic} two-body problem as a ``post-Minkowskian" (PM) expansion in powers of $G_N$, one would have to compute these Feynman diagrams for general worldlines $x_{a=1,2}^\mu(\tau)$ with a given set of initial conditions.   At present, this is only tractable for scattering (unbound) trajectories\footnote{A formalism for mapping scattering (PM) to bound state (PN) observables has been recently proposed in refs.~\cite{Kalin:2019rwq,Kalin:2019inp,Cho:2021arx}.} at large impact parameter $G_N E_{cm}/b\ll 1$.    In this case, the structure of the resulting loop momentum integrals is sufficiently well understood, and, when defined via dimensional regularization, can be treated by techniques already developed for perturbative quantum field theory calculations in high energy physics~\cite{Weinzierl:2022eaz}.  Conservative and radiative dynamics in PM scattering has been an active area of recent effort that has brought together theoretical approaches from diverse communities, including researchers working on scattering amplitudes and formal theory, general relativity, and on effective field theories.    See the Snowmass whitepaper~\cite{Buonanno:2022pgc} for a review and a complete set of references.

   Various PM effects in relativistic two-body scattering, treating classical gravity by the sort of EFT ideas reviewed here, have been recently studied in refs.~\cite{Goldberger:2016iau,Goldberger:2017ogt,Li:2018qap,Goldberger:2019xef,Kalin:2020mvi,Goldberger:2020wbx,Kalin:2020fhe,Kalin:2020lmz,Mougiakakos:2021ckm,Liu:2021zxr,Dlapa:2021npj,Riva:2021vnj,Dlapa:2021vgp,Mougiakakos:2022sic}.    In the case of relativistic hard graviton scattering in flat spacetime,   soft and collinear singularities have been treated by EFT methods in~\cite{Beneke:2012xa,Okui:2017all,Chakraborty:2019tem,Beneke:2021aip}.      For bound orbits, relevant to the LIGO problem, the perturbative expansion involves simultaneously powers of $G_N$ and of $v^2\sim G_N M/r$ as described above.   We turn to the EFT formulation of such non-relativistic bound states in the next section.

\subsection{NRGR and \zosoir}
\label{sec:NRGR}

The formalism outlined so far is suitable for widely separated but relativistic compact objects, in the kinematic regime $G_N E/b\ll 1$ .   In this limit  the Feynman rules of the EFT provide a systematic double expansion in powers of $G_N E/b\ll 1$ as well as powers of $\hbar /L\ll 1$, where $L\sim E b$ is the orbital angular momentum scale of the binary.    In particular, the only physical scale appearing in any Feynman integral over internal graviton momenta is the separation $b$, so the effective theory has manifest powers counting in $G_N E/b \ll 1$, with $\hbar /L\ll 1$ serving to count the number of internal graviton loops in a given graph.  Objects that move relativistically but interact weakly through graviton exchange cannot form bound orbits, so the formalism as it stands is suitable to treat PM scattering kinematics.   

The Lorentz covariant Feynman rules associated with the diagrammatic expansion of Eq.~(\ref{eq:inin}) are not yet optimized to do calculations in the PN expansion.   For $v\ll 1$ there is now a hierarchy of scales between orbital dynamics at distances $\sim r$ and radiation emission at $r/v\gg r$.    If the Feynman integrals are defined by dimensional regularization, there are two regions of internal graviton momenta where the integrals are supported:
\begin{itemize}
\item Potential:   $(p^0,{\vec p})\sim (v/r,1/r)$.
\item Radiation:   $(k^0,{\vec k})\sim (v/r,v/r)$.
\end{itemize}
The potential region  corresponds to off-shell gravitons which are exchanged between the compact objects.   In position space, they generate instantaneous in time, long range range forces between the particles, binding them into quasi-elliptical orbits.  Radiation gravitons can go on-shell, propagating out to the detector, or remain off-shell, generating both ``dissipative" (time reversal odd) and ``conservative" ($T$-even) radiation reaction forces.    In dimensional regularization, a given Feynman integral can be ``threshold expanded"~\cite{Beneke:1997zp} around the various configurations of potential and radiation regions that follow from the kinematics (method of regions).   The expanded Feynman integral is then equivalent to a linear combination of simpler integrals, each containing a single physical scale.   These simplified integrals can now be calculated for arbitrary (bound or unbound) non-relativistic trajectories ${\vec x}_{1,2}(t)$, as is necessary for the inspiral problem, and scale homogeneously as definite powers of the expansion parameter $v$.

Rather than expanding out the PM Feynman diagrams in powers of $v$, it is more efficient to perform the expansion at the level of the action, by explicitly decomposing the graviton field into modes with support around the potential and radiation region,
$$
h_{\mu\nu}(x)={\hat h}_{\mu\nu}(x) + \int {d^3{\vec p} \over (2\pi)^3}e^{i{\vec p}\cdot {\vec x}}  H_{\mu\nu;\bf p}(x^0),
$$
where we assign the scaling $x^\mu\sim r/v$, ${\vec p}\sim 1/r$, and therefore spacetime derivatives acting on radiation ${\hat h}_{\mu\nu}$ and potential $H_{\mu\nu;\bf p}(x^0),$ fields scale uniformly, $\partial_\mu\sim v/r$.   Because the kinetic term for the $H_{\mu\nu;\bf p}(x^0)$ field is suppressed relative to spatial gradient energy, it is a non-propagating mode with instantaneous in time propagator.   In addition, it is necessary to perform a multipole decomposition of  ${\hat h}_{\mu\nu}$.   \eg~, choosing the center of mass frame,
$$
{\vec X}_{cm} ={\int d^3 {\vec x} \tau^{00} {\vec x}\over \int d^3 {\vec x} \tau^{00}}\equiv 0,
$$
we replace
$$
{\hat h}_{\mu\nu}(x) \mapsto \sum_n {\vec x}^{i_1}\cdots   {\vec x}^{i_n} \partial_{i_1}\cdots \partial_{i_n} {\hat h }_{\mu\nu}(x^0,0),
$$
at the level of the terms in the Lagrangian involving couplings to potential modes or to the non-relativistic worldlines.

After performing these field redefinitions, we have an EFT of radiation and potential gravitons coupled to non-relativistic particles, whose Feynman rules scale as definite powers of $L\sim m v r$ and $v$:
$$
\begin{array}{c|c|c|c|c|c}
\partial_\mu & {\vec x}_a & {\vec p} & {\hat h} & H_{\vec p}  & m/m_{Pk}\\
\hline
v/r& r & 1/r & v/r & r^2 \sqrt{v} & \sqrt{L v}
\end{array}
$$
For example, the contributions to the effective action of Eq.~(\ref{eq:inin}) from diagrams with $n$ external (background) gravitons and no internal graviton loops scale as $L^{1-n/2}$ times powers of $v$, whereas graviton loops are suppressed by powers of $1/L$ in the classical limit $L\sim m v r\gg\hbar$.   Therefore, in the classical limit, only diagrams with at most one external graviton, $n\leq 1$, are relevant.   We refer to the theory of potentials and radiation by the acronym NRGR~\cite{Goldberger:2004jt} to emphasize the analogy to NRQCD~\cite{Caswell:1985ui}, the EFT description of on-relativistic bound states $Q{\bar Q}$  ($M_Q\gg \Lambda_{QCD}$), which employs a similar mode decomposition~\cite{Luke:1999kz} and multipole expansion~\cite{Grinstein:1997gv} of the gluon fields in full QCD.

Because the potential gravitons cannot go on-shell, it is possible to integrate them out to obtain a local EFT of self-interacting radiation gravitons coupled to the bound state.    We refer to this EFT, valid at distances longer than the orbital scale $r$, as ``\zosoir'' because it encodes the interactions of a \emph{${\cal Z}$oomed ${\cal O}$ut ${\cal S}$ingle ${\cal O}$bject} whose internal structure (the binary constituents) cannot be directly resolved by long wavelength radiation modes.    The composite object is defined in terms of a worldline variable that tracks the motion of the center of mass, an orthonormal frame that accounts for the spatial orientation relative to asymptotic inertial observers, and finally an infinite tower of electric and magnetic ``mass" and ``current" multipoles respectively.

If one chooses a gauge for the potential gravitons that preserves long wavelength diffs acting on the radiation field, the form of the Lagrangian for the composite defect coupled to gravity takes the general form~\cite{Goldberger:2005cd,Goldberger:2009qd}
\begin{equation}
\label{eq:zoso}
{\cal S}_{\includegraphics[width=0.05\hsize,valign=c]{zosoir}} = -\int d\tau L(X) -\int dx^\mu S_{ab} \omega_\mu^{ab} + {1\over 2} \int d\tau I_{ab} E^{ab} + {1\over 2} \int d\tau J_{ab} B^{ab} +{1\over 6}\int d\tau_{abc} \nabla^c E^{ab}+\cdots 
\end{equation}
in the center of mass frame, ${\vec X}_{cm}=0$, ${\vec P}^i_{CM}=\int d^3{\vec x} \tau^{0i}=0$.   The Lagrangian consists of an infinite tower of multipole moments $I_{a_1\cdot a_\ell}$, $J_{a_1\cdot a_\ell}$ of parity $(-1)^\ell$, $(-1)^{\ell+1}$ respectively, constructed out of the positions and the spins of the binary constituents.    These moments couple linearly to the gradients of the Weyl curvature, and source emission of radiation out to the detector.  They can be interpreted as Wilson coefficients for operators in the EFT that results from integrating out the binary dynamics at short distance scales of order $r$.   As in other EFTs of gravity (\eg~ in cosmology~\cite{Cheung:2007st,Weinberg:2008hq,Carrasco:2012cv,Porto:2013qua}), the Wilson coefficients have non-trivial time dependence, due to the breaking of Poincare invariance by sources.    The full Lagrangian below frequencies $1/r$ is then the sum of ${\cal S}_{\includegraphics[width=0.05\hsize,valign=c]{zosoir}}$ and the Einstein-Hilbert term for $g_{\mu\nu}=\eta_{\mu\nu}+{\hat h}_{\mu\nu}/m_{Pl}$.    For a more detailed explanation of the notation in Eq.~(\ref{eq:zoso}) , see \eg~ ref.~\cite{Goldberger:2009qd}.

\begin{figure}
\begin{center}
\includegraphics[width=0.50\hsize]{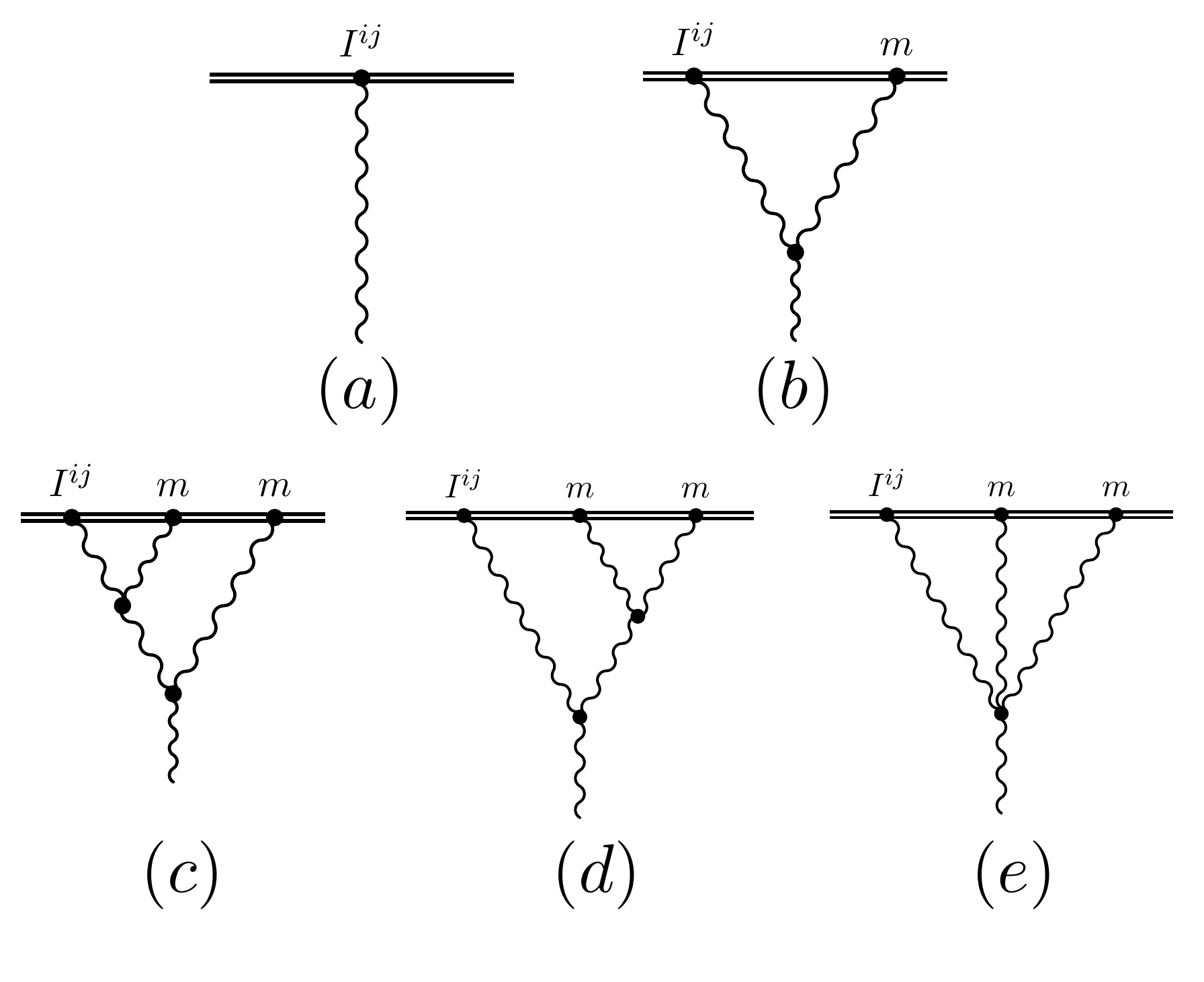}
\end{center}
\caption{Leading order quadrupole emission (a) in \zosoir~and perturbative ``tail'' corrections at ${\cal O}(G_N E\omega)$ (b) and ${\cal O}(G_N E\omega)^2$ (c)-(e).   Diagram (b) has a $1/\epsilon_{IR}$ singularity in dimensional regularization, while (c)-(e) contain both IR and UV poles in $d=4-\epsilon$ spacetime dimensions.}
\label{fig:tails}
\end{figure}

The Lagrangian~(\ref{eq:zoso}) is actually universal, in the sense that it describes soft graviton emission from a completely generic self-gravitating system of characteristic size $\sim b$ and ADM energy $\sim E$.   The moments then scale as $\sim E b^\ell$ and the EFT has manifest power counting in the multipole expansion parameter  $\omega b\ll 1$ a well as the quantity $\hbar/E b\ll 1$ that controls quantum corrections.    The EFT is valid regardless of wether $G_N E/b \ll 1$ or not.   In the latter case, the insertions of the mass ($\ell=0$) moment are non-perturbative and must be re-summed.   This is formally equivalent to expanding the Lagrangian about a fixed non-trivial background ${\bar g}_{\mu\nu}$ corresponding to the Kerr metric with mass $E$ and spin $S^{ab}$, and using curved spacetime propagators and vertices for the radiation graviton Feynman rules. 

 In the opposite limit, the EFT is a double expansion in $\omega b\ll 1$ and  $G_N E/b\ll 1$, in which case the graviton propagates in flat spacetime and insertions of the mass monopole into Feynman diagrams can be treated perturbatively.   In the classical limit $\hbar/E b\ll 1$ only the diagrams with a single external graviton survive, see Fig.~\ref{fig:tails}.   The resulting Feynman integrals are tractable by standard techniques~\cite{Weinzierl:2022eaz} and, at least at sufficiently small orders in $G_N E/b\ll 1$ are calculable analytically for arbitrary time-dependent source moments $I_{a_1\cdots a_\ell}$, $J_{a_1\cdots a_\ell}$.

Focusing on non-relativistic binaries, we assign power counting $I_{a_1\cdots a_\ell}\sim M r^\ell$, $J_{a_1\cdots a_\ell}\sim M v r^\ell$, in which case the two expansion parameters of the EFT control effects down by different powers in $v$:   multipole corrections in powers of $\omega r\sim v$ and gravitational wave ``tails" (perturbative scattering of outgoing radiation off the total mass $M$ of the binary) in powers of $G_N M/r\sim v^3$.   In this regime, it is possible to obtain the Wilson coefficients in Eq.~(\ref{eq:zoso}) by matching to NRGR at distance scales $\gsim r$ where the two theories are both valid.   For applications to classical binary inspirals it is only necessary to match to NRGR in the sectors with zero or one external radiation gravitons.  Using diffeomorphism invariance, this is sufficient to determine the relevant non-linear couplings of the radiation mode as well.

Matching to NRGR, the terms with zero external gravitons determine the conservative part of the two-body interaction Lagrangian induced by potential graviton exchange.   For the calculation of such potentials, the relevant momentum space Feynman integrals are equivalent to those that one would encounter in the calculation of a two-point function in a massless Euclidean quantum field theory in $3-\epsilon$ spatial dimensions~\cite{Goldberger:2004jt}.    In a generic gauge,  the potentials at the $n$PN order require knowledge of $n$-loop Feynman integrals.    However, by exploiting a convenient field redefinition of the graviton that is well suited to the non-relativistic limit, introduced in refs.~\cite{Kol:2007rx,Kol:2007bc,Kol:2010ze}, it is possible to postpone the number of loops by one order in perturbation theory.   Within the EFT approach, the non-relativistic spin-independent potentials at 2PN order where computed in ref.~\cite{Gilmore:2008gq}, which introduced some of the tools necessary to carry out higher order PN loop diagrams.     The systematic study of higher order spinless PN potentials was initiated in~\cite{Foffa:2011ub} and extended~\cite{Foffa:2012rn,Foffa:2013gja,Foffa:2016rgu} to 4PN in~\cite{Foffa:2016rgu}.      A complementary approach to the computation of gravitational potentials at order 4PN and beyond can be found in refs.~\cite{Blumlein:2019zku,Blumlein:2020pog,Blumlein:2020znm,Blumlein:2020pyo,Blumlein:2021txj,Blumlein:2021txe}.

For phenomenological applications, it is necessary also to incorporate spin-dependent potentials into the PN predictions.   The methodology for computing such effects within the EFT was introduced in~\cite{Porto:2005ac}.   It has been used in the computation of spin-spin (``hyperfine")~\cite{Porto:2006bt,Porto:2007tt,Porto:2008tb,Levi:2008nh,Levi:2011eq,Levi:2014sba}  gravitational interactions,  spin squared~\cite{Porto:2008jj,Levi:2014gsa,Levi:2015ixa,Levi:2016ofk,Levi:2020uwu,Kim:2021rfj} effects associated with finite size, higher PN spin-orbit coupling effects~\cite{Porto:2007px,Porto:2010tr,Levi:2010zu,Levi:2015uxa,Levi:2020kvb,Cho:2022syn} and even spin effects beyond quadratic order~\cite{Levi:2015msa,Levi:2019kgk,Levi:2020lfn}.

Matching to the single soft graviton emission amplitude in NRGR determines the multipole moments as functions of the particle orbital and spin variables.    This was carried out to 1PN order in~\cite{Goldberger:2009qd} in the case of spinless binaries, and extended to 2PN order in ref.~\cite{Leibovich:2019cxo}.    The relation between the energy momentum pseudotensor and the moments as defined in Eq.~(\ref{eq:zoso}) was made systematic to all orders in the multipole expansion in ref.~\cite{Ross:2012fc}.  Spin corrections to the multipole moments have been obtained refs. in~\cite{Porto:2010zg,Porto:2012as,Pardo:2020hxc,Cho:2022syn} where the current state of the art~\cite{Cho:2022syn} includes spin effects up to 4PN order.

It is convenient to compute radiative corrections to binary dynamics , \eg~effects from radiation graviton exchange, directly in the radiation EFT of Eq.~(\ref{eq:zoso}) rather than in NRGR.   This has the advantage that the results are universal, \ie~can be obtained without knowledge of the explicit form of the multipoles, and therefore are valid for describing soft graviton radiation from an arbitrary energy-momentum distribution of finite extent.     For astrophysical applications, the relevant quantities are the zero point function, which encodes the radiative corrections to the equations of motion (radiation reaction forces), and the one-point function, which determines the waveform measured at the detector as a function of the time-dependent moments evaluated on the solutions to PN equations of motion.

In such calculations, one encounters both UV and IR logarithmically divergent Feynman diagrams\footnote{The resulting Feynman integrals take the form of 3D integrals over Euclidean loop momentum involving ``massive" propagators $1/({\vec \ell}^2-\omega^2)$ at fixed (complex) values of the external frequencies $\omega$.   Restricting the EFT to the sector with at most one external  radiation graviton implies that at most one external momentum can show up in the propagators.}.  In order to preserve manifest diff invariance, these are defined via dimensional regularization, where the log divergences correspond to poles in $\epsilon=4-d$.   The IR divergences arise from so-called ``gravitational wave tails,'' which refers to the distortion of the outgoing graviton wavefunctions in the $1/r$ gravitational potential sourced by the mass monopole, as depicted in Fig~\ref{fig:tails}(b)-(e).     They are analogous to the IR divergences found in non-relativistic Coulomb scattering, and in the gravitational context appear first at order $G_N M\omega\sim v^3$ beyond leading order, and then again at every subsequent order in power of  $G_N M\omega$.   The resolution~\cite{Goldberger:2009qd,Porto:2012as} of these IR divergences is similar to the QED case~\cite{Weinberg:1965nx}:    in frequency space,  the logarithmic dependence on the IR regulator exponentiates to all orders in $G_N M\omega$ into an overall phase factor multiplying the graviton emission amplitude~\cite{Goldberger:2009qd}.   This phase then cancels in the gravitational energy flux (emitted power), which depends on the modulus squared of the amplitude.   Similarly, upon transforming the amplitude to the time domain, the IR divergent phase simply has the effect of shifting the argument of the gravitational wave signal $h(t)$ recorded at the detector.  This shift is arbitrary, and is absorbed into the definition of the (experimentally determined) ``initial time'' when the signal first enters the detector's frequency band~\cite{Porto:2012as}.    Even though the dependence on the IR regulator disappears from infrared safe physical observables, the gravitational wave tails leave a measurable imprint on the waveform $h(t)$:   in the frequency domain, the graviton emission amplitude squared factorizes as $S(\omega)\times |{\cal A}(\omega)|^2$~\cite{Asada:1997zu,Khriplovich:1997ms}, where $S(\omega)=4\pi G_N M\omega/(1-e^{-4\pi G_N M \omega})$, is the Sommerfeld factor familiar from Coulomb scattering in non-relativistic quantum mechanics.

The UV logarithmic divergences also have a a standard quantum field theoretic resolution~\cite{Goldberger:2009qd}.   In \zosoir, they correspond to singularities due to graviton propagation in the short distance part of the source's gravitational potential, induced by relativistic corrections involving higher powers of $G_N M/r$.   As such, they correspond to singularities in the multipole expansion, which are resolved by the finite orbital separation between the binary constituents. They appear for instance in the graviton emission amplitude (one-point function) at order $(G_N M\omega)^2\sim v^6$, or 3PN,  relative to the leading order emission, see Fig.~\ref{fig:tails}(c)-(e).

The UV divergent terms are analytic in the frequency, and can be absorbed into the Wilson coefficients in the \zosoir~ Lagrangian.   As such, they imply the existence of RG equations for the suitably renormalized multipole moments.   For instance, the RG equation for the $\ell=2$ mass moment in frequency space is given by~\cite{Goldberger:2009qd}
$$
\mu {d\over d\mu} I_{ab}(\omega,\mu) = -{214\over 105} (G_N M\omega)^2 I_{ab}(\omega,\mu),
$$
which is a universal result that holds for arbitrary systems that emit soft radiation with $G_N M\omega\ll 1$.   In the binary problem, running the RG from a scale $\mu_0\sim r$ of order the orbital scale in the UV to a scale $\mu\sim \omega\sim v/r$ in the IR, then predicts the entire pattern of logarithms of velocity in the mass quadrupole channel.  \eg~, for a binary system in a circular orbit with angular frequency $\omega$, one finds
$$
{{\dot E}^{\ell=2_E}_{\mbox{log}}\over {\dot E}^{\ell=2_E}_{\mbox{LO}}}=\left[\mu\over \mu_0\right]^{-{428\over 105} (G_N M\omega)^2}=1-{428\over 105} v^6 \ln v + {91592\over 11025} v^{12} \ln^2 v -{39201376\over 347287} v^{18}\ln^3 v +\cdots.
$$
Similarly RG flows occur at higher PN orders for the $\ell=0$ moment $M$~\cite{Goldberger:2012kf,Galley:2015kus}, and for generally for the mass and current multipoles at each $\ell\geq 2$~\cite{Almeida:2021jyt}.   Of course, the RG by itself does not fix the UV scale $\mu_0\sim 1/r$ where we define the Wilson coefficients.   That must be determined by doing a 3PN matching calculation  to NRGR, where the orbital scale is non-zero and the UV divergent logs of \zosoir~get replaced by finite logarithms of the orbital separation.

The EFT defined by Eq.~(\ref{eq:zoso}) is also well suited to the calculation of radiative corrections to the binary equations of motion induced by gravitational wave emission.    Classical radiation reaction in worldline EFTs was first studied in the context of finite size corrections to the Abraham-Lorentz-Dirac equations of motion in classical electrodynamics in ref.~\cite{Galley:2010es}.    The definitive approach to radiation reaction in the gravitational case, using the full Schwinger-Keldysh machinery, was initiated  by Galley and Tiglio~\cite{Galley:2009px}, and has been extended to ultra-relativistic sources in ref.~\cite{Galley:2013eba}, to include higher order PN corrections, \eg~ tails and other ``hereditary'' or ``memory" effects, in  refs.~\cite{Foffa:2011np,Galley:2012qs,Goldberger:2012kf,Galley:2015kus,Foffa:2019eeb,Foffa:2021pkg,Almeida:2021xwn,Edison:2022cdu} and to include the spin of the binary constituents in refs.~\cite{Maia:2017gxn,Maia:2017yok,Pardo:2020hxc,Cho:2021mqw}.    We note that, in general, radiation reaction effects need not be purely dissipative.  For example, at 4PN order, there can be $T$-even conservative contributions to the equations of motion due to radiation.   Such conservative contributions have been studied in detail in refs.~\cite{Galley:2015kus,Porto:2017dgs,Porto:2017shd,Foffa:2019rdf,Foffa:2019yfl}, which explain the precise way in which the radiation and potential sectors of the EFT conspire to cancel unphysical IR divergences that arise at intermediate steps, yielding unambiguous IR finite predictions for binary dynamics at 4PN order.

\subsection{\zosouv}
\label{sec:UV}

The formalism as described so far is adequate for compact objects whose internal structure is gapped, so that gravitational interactions at scales longer than the orbital radius cannot irreversibly modify the intrinsic properties of the object.   For black holes, though, this frequency gap is of order $1/r_s$, so that finite size effects will come in at some finite order in $r_s/r\sim v^2$ in the PN expansion.   In order to have a fully systematic treatment of PN black hole binary dynamics, such dissipative effects cannot be neglected.

For instance, from black hole perturbation theory~\cite{Sasaki:2003xr} it is known that the change in mass due to tidal heating induced by a small binary companion in a bound orbit appears at order 4PN~\cite{Poisson:1994yf} in the Schwarzschild case, and becomes enhanced to 2.5PN~\cite{Tagoshi:1997jy} for (near extremal) Kerr black holes.    In the latter case, the tidal interactions can actually \emph{decrease}~\cite{DEath:1975jps,Poisson:2004cw} the mass of the black hole as a consequence of stimulated emission (``rotational superradiance")~\cite{zeldovich,staro}, a field theoretic realization of the Penrose process~\cite{Penrose:1971uk} of energy extraction from the black hole's ergosphere.  

On general grounds~\cite{Callen:1951vq}, dissipation, \eg~ flux of energy and angular momentum across the surface of the compact star, signals the presence of a continuum spectrum of localized degrees of freedom that couple to gravity in the bulk spacetime.  For a neutron star, the additional degrees of freedom correspond to the low-lying hydrodynamic modes of nuclear matter, while for classical black holes, the horizon fluctuations are presumably related to the quasinormal mode solutions of the Teukolsky equation.   Regardless of the microscopic origin of the internal degrees of freedom, their presence has an effect on the binary inspiral dynamics at some order in the PN expansion.    

It is therefore useful to have a way of incorporating the effects of dissipation directly in the worldline description of the compact objects, without explicitly having to track the evolution of the internal modes themselves.    An EFT framework for this was first introduced in ref.~\cite{Goldberger:2005cd}, which describes the long wavelength dissipative response of compact objects to external gravitational perturbations, by ``integrating in''  a \emph{quantum mechanical} $0+1$-dimensional defect field theory of degrees of freedom localized on the worldline.

Independent of their UV origin, in the long distance limit these modes have local diff and reparameterization invariant couplings to the spacetime curvature.   Organizing the algebra of defect operators in terms of the linearly realized rotations about the objects spatial location, the symmetries of the EFT guarantee that the Lagrangian must be of identical form to Eq.~(\ref{eq:zoso}), where now the multipole moments $I_{a_1\cdots a_\ell},J_{a_1\cdots a_\ell},$ should be regarded as a set of composite operators constructed out of the microscopic degrees of freedom, acting on some internal Hilbert space of physical states.    As long as we probe the system with slowly varying fields, the compact object itself also appears as a \emph{${\cal Z}$oomed ${\cal O}$ut ${\cal S}$ingle ${\cal O}$bject}, even if its internal structure is arbitrarily complicated.

We do not need to know what the internal modes are in order to make  predictions in the infrared.   In this case, long distance observables can be calculated in terms of the correlation functions of the multipole operators, which in turn are determined by a matching calculation to the UV theory.   The power counting of the EFT indicates that at long distances, the leading contribution is from the two-point correlators of the electric and magnetic quadrupole operators
$$
\langle  I_{ab}(\tau) I_{cd}(0)\rangle, \langle  J_{ab}(\tau) J_{cd}(0)\rangle
$$
evaluated in the equilibrium (pure or mixed) state of the object.   Predictions in the EFT, in powers of $\omega {\cal R}\ll 1$ are systematically improvable by including more multipoles, higher-point correlators, or perturbative graviton interactions which scale as powers of  $G_N M\omega\lsim \omega{\cal R}\ll 1$.  

To match to this \zosouv~we compute on-shell graviton scattering off an isolated object in the EFT, using Eq.~(\ref{eq:zoso}), and compare it to the low frequency limit of the corresponding observable in the full theory.  As an example, consider graviton absorption by a Schwarzschild black hole, which in the EFT has the matrix element
$$
i{\cal A}(M\rightarrow X) \approx {i\over 2 m_{Pl}}\int dt e^{-i\omega t} \langle X|I_{ab}(t) |M\rangle\times \langle 0|E_{ab}(t,0)|k,h\rangle +\mbox{magnetic}
$$
in the rest frame, to leading order in $1/m_{Pl}$.  Here, the matrix element $\langle 0|E_{ab}(t,0)|k,h\rangle$ between the one-graviton state of four-momentum $k^\mu$ ($k^0=\omega>0)$, helicity $h=\pm 2$, and the vacuum is readily computed by standard canonical quantization of Eq.~(\ref{eq:eh}), as in~\cite{Gupta:1952zz,Feynman:1963ax,DeWitt:1967yk,DeWitt:1967ub,DeWitt:1967uc}.   The transition matrix element $ \langle X|I_{ab}(t) |M\rangle$  from the initial black hole of mass $M$ to some unknown final state $|X\rangle$ is not calculable in the EFT, but asuming unitarity
$$
\sum_X |X\rangle \langle X| = \mathbf{1}_{\cal H},
$$
we can express the inclusive absorption cross section for a graviton incident on the horizon,
$$
\sigma_{abs}(\omega) = \lim_{T\rightarrow\infty} {1\over 2\omega}\sum_X {|{\cal A}(M\rightarrow X)|^2 \over T}=G_N\pi \omega^3 \int dt e^{i\omega t} \epsilon^{*}_{cd,h}(k) \langle I_{cd}(t) I_{ab}(0)\rangle  \epsilon_{ab,h}(k) + \mbox{magnetic},
$$
in terms of the two-point correlators of the $\ell=2$ multipole operators in the initial state of a black hole of mass $M$ and spin $J$.    

The cross section $\sigma_{abs}(\omega)$ is a physical quantity that can be compared against the predictions of classical general relativity in the limit $r_s\omega\ll 1$, where the EFT description is useful.  Using the classical absorption probabilities calculated in refs.~\cite{staro,Page:1976df} one finds that $\sigma_{abs}(\omega)\approx 4\pi r_s^6\omega^4/45$, and exploiting the rotational invariance of the Schwarzschild black hole to write
$$
\int dt e^{i\omega t} \langle I_{ab}(t) I_{cd}(0)\rangle ={1\over 2}\left[\eta^\perp_{ac}\eta^\perp_{bd}+\eta^\perp_{ad}\eta^\perp_{bc}-{2\over 3} \eta^\perp_{ab}\eta^\perp_{cd}\right] A_+^E(\omega) ,
$$
($\eta^\perp_{ab}=\eta_{ab} - p_a p_b/M^2$ is the spatial metric in the black hole's rest frame) one finds that the frequency space correlators are $A^E_+(\omega)=A^B_+(\omega)\approx 2 \theta(\omega) r_s^6\omega/45G_N$ to leading order in $r_s\omega\ll 1$.   In the case of spinning objects, the correlator is no longer determined by a single form factor $A^{E,B}_+(\omega)$ as more tensor structures, involving the spin vector of the object, can appear.   For non-zero spin, the EFT has been extended to slowly spinning black holes~\cite{Porto:2007qi} , to more general spinning sources in ref.~\cite{Endlich:2016jgc}, and generalized to rapidly spinning (close to extremal) Kerr black holes~\cite{Goldberger:2020fot}.

 The point of this exercise is that the same\footnote{The correlators that appear in classical binary evolution are the retarded (causal) Green's functions $-i\theta(\tau) \langle [{\cal O}(\tau),{\cal O}(0)]$ rather than the Wightman functions.    The causal two-point correlator is related to the Wightman function by a dispersion relation whose form in~\zosouv~was worked out in the non-spinning case in~\cite{Chakrabarti:2013lua,Goldberger:2020wbx} and in ~\cite{Goldberger:2020fot} for non-zero spin.}  correlators that one extracts from on-shell observables  in the one-body sector also control \emph{off-shell} graviton exchange processes in the two-body sector, where the binary dynamics is described by NRGR.  By including diagrams with insertions of the multipole operators $I_{ab}$, $J_{ab}$, it becomes possible to include the effects of horizon dynamics in the EFT while retaining a worldline description of the binary constituents.  For example, single graviton exchange between two black holes generates a tidal friction ($T$-odd) term in the two-body equations of motion associated with the excitation of horizon modes, leading to a flux of energy across the event horizon of the form 
 \begin{equation}
 \label{eq:absEdot}
\left. {dE\over dt}\right|_{h}={8\over 5} {G_N^5 m_1^2 m_2\over |{\vec x}_1-{\vec x}_2|} (m_1+m_2)\left[1+3 \chi_1^2 -{15\over 4}\chi_1^2\left({{\vec s}_1\cdot ({\vec x_1}-{\vec x}_2)\over |{\vec x}_1-{\vec x}_2|}\right)^2\right] {\vec S}_1\cdot {\vec L}+(1\leftrightarrow 2),
 \end{equation}
 at leading PN order.

 For nearly maximally rotating black holes, with $\chi=|{\vec S}|/G_N M\lsim 1,$  Eq.~(\ref{eq:absEdot}) gets enhanced, by the superradiant effect, to 2.5PN order compared to 4PN in the Schwarzschild case, $\chi=0$.   Notice that the energy flux can have either sign depending on the relative orientation between the black hole spin and the orbital angular momentum ${\vec L}$.   In particular, it is possible to extract rotational energy from the black holes as in the Penrose process~\cite{Penrose:1971uk}. Eq.~(\ref{eq:absEdot}) generalizes to arbitrary orbits and spin orientations earlier results obtained by classical techniques in~\cite{Alvi:2001mx,Chatziioannou:2012gq,Chatziioannou:2016kem}.    Other applications of \zosouv~to the tidal interactions of compact objects can be found in refs.~\cite{Chakrabarti:2013lua,Chakrabarti:2013xza,Endlich:2015mke,Steinhoff:2016rfi,Baumann:2018vus,Wong:2019yoc,Wong:2019kru,Baumann:2019ztm,Wong:2020qom,Gupta:2020lnv,Martinez:2020loq,Creci:2021rkz,Martinez:2021mkl}.   An analogous formalism for dissipative effects in cosmological EFTs was introduced in~\cite{LopezNacir:2011kk}.

Because the effects of the horizon become as large as  2.5PN order, it is phenomenologically well motivated to study corrections to this result.    These have been obtained in the extreme mass limit using black hole perturbation theory in ref.~\cite{Tagoshi:1997jy}, and in refs.~\cite{Chatziioannou:2012gq,Chatziioannou:2016kem} as a specific case of a more general process where the black hole propagates in a background spacetime whose curvature scale is large.   In the regime where these two distinct expansion schemes overlap, the 1PN limit of the results in~\cite{Tagoshi:1997jy,Chatziioannou:2012gq} for the horizon energy and angular momentum fluxes are in agreement.   However, there is a discrepancy at order 1.5PN between the more general methods used in~\cite{Chatziioannou:2016kem} and those of~\cite{Tagoshi:1997jy} obtained by analytically solving the Teukolsky equation in the $r_s\omega \ll 1$ PN limit.  This disagreement motivates an independent computation of these effects, using the EFT techniques outlined here, which is underway.

\subsection{Quantum effects}

Because the \zosouv~formalism is explictly quantum mechanical, it is capable of describing processes involving quantum black holes interacting with other particles or fields.   In particular, it can be used~\cite{Goldberger:2019sya,Goldberger:2020geb} to take into account the effects of Hawking radiation on scattering observables.   The EFT description assumes that black holes behave according to quantum mechanical rules, with unitary time evolution and a complete Hilbert space of microstates.    It is valid in the window of momentum transfers $q$ defined by
$$
t^{-1}_{Page}\ll q\ll 1/G_N M\ll m_{Pl},
$$
where, \eg~ in the Schwarzschild case, the black hole evaporation scale is of order the Page time~\cite{Page:1976df}, $t_{Page} \sim G_N^2 M_{BH}^3$.   The upper bound ensures that the worldline description of the black hole is reliable, while the lower bound allows us to treat black holes as approximately long-lived asymptotic states in the $S$-matrix.   Since $G_N M_{BH}\gg 1/m_{PL}$, the black holes can be regarded as being semiclassical.    Because the time scales in a given process are short compared to  $t_{Page}$,  we need not worry about having to take into account whatever physics is responsible for unitarizing the evaporation process, see \eg~\cite{Almheiri:2020cfm} for a summary of current opinion on this subject.

To construct the EFT, we have to determine how the structure of multipole operator correlation functions is modified by the presence of a Hawking thermal spectrum of radiation emission from the black hole.    As in the classical case, we match to the simplest possible observables that depend, on the EFT side, on the multipole correlators.    It is convenient in particular to calculate the on-shell transition probabilities $p(n\rightarrow m)$ for a Kerr black hole to emit $m$ gravitons out to future null infinity, given that $n$ particles each of fixed energy $\omega$ are incident on the black hole in the far past.    Explicit results~\cite{Bekenstein:1977mv,Panangaden:1977pc}  are available\footnote{The results of refs.~\cite{Bekenstein:1977mv,Panangaden:1977pc} are for free scalar fields propagating in the black hole background, but they generalize naturally to higher spin $s>0$ fields by simply replacing the scalar transmission coefficients by their higher spin version found in refs.~\cite{staro,Page:1976df}.}  for this observable, in the limit of free quantum field theory in the black hole background.

In \zosouv, one finds that to leading order in $\omega/m_{Pl}\ll 1$, the $n\rightarrow m$-particle probabilities are controlled by $n+m$-point Wightman correlation functions of the multipole operators in Eq.~(\ref{eq:zoso}).  Given the structure of the full theory result~\cite{Bekenstein:1977mv,Panangaden:1977pc}, these higher points correlators factorize into suitable products of two-point functions, up to non-Gaussianities suppressed by $\omega^2/m^2_{Pl}\ll 1$.    Somewhat surprisingly, one finds that at $r_s\omega \ll 1$ the effects of Hawking radiation are not Planck suppressed at the level of the Wightman functions.   Instead, they become \emph{enhanced}  in the limit $\hbar\omega/T_H= 4\pi r_s\omega \ll 1$ where the EFT is valid, a consequence of the high temperature behavior of the Planck distribution.

Despite this enhancement at the level of the Wightman functions, the effects of Hawking radiation cancel at the level of free field \emph{retarded} correlators.  Up to corrections suppressed by $1/m_{Pl}$ these take the same form in the  Unruh state~\cite{Unruh:1976db} that describes an evaporating black hole or the Boulware state~\cite{Boulware:1974dm} where the black hole does not emit radiation.    Consequently,  finite size effects associated with the quantized nature of the black hole horizon, \eg~ in the observables discussed in sec~\ref{sec:UV}, are suppressed by at least one power of $\omega^2/m_{PL}^2\ll 1$.  Thus, if one assumes that black holes evolve according to the usual rules of quantum mechanics, the results of~\cite{Goldberger:2019sya} imply a no-go theorem on the possibility of detecting black hole hair or other possible exotic signatures of quantum behavior in  binary black hole mergers at LIGO/VIRGO or any other foreseeable experiment.

On the other hand, the emission of Hawking radiation does modify observables that depend on the Wightman functions directly.   While not of phenomenological important, an example~\cite{Goldberger:2020geb} of formal interest is the inelastic scattering of elementary particles incident on a semiclassical black hole, mediated by the exchange of \emph{virtual} (off-shell) Hawking gravitons.  For illustration, consider a scalar particle $\phi$ with mass in the range $k_B T_H\ll m_\phi \ll M_{BH}$.   In this window, direct $s$-channel product of $\phi$-particle Hawking pairs is exponentially (Boltzmann) suppressed, so that the scattering process
$$
\phi(p)+\mbox{BH}\rightarrow \phi(p')+{\mbox{BH}}^\prime
$$
proceeds instead through graviton exchange.   One finds the result
\begin{eqnarray}
\nonumber
{d^3\sigma \over dq^2 d(q\cdot v)} &\approx& {7 G_N  r_s^5\over 270\pi [(v\cdot p)^2 - m_\phi^2]} \left[(v\cdot p)^4  -m_\phi^2 (v\cdot p)^2\left(1 - {12\over 7} {(v\cdot q)^2\over q^2}\right) \right.\\
\label{eq:result}
& & \left.+{1\over 7}m^4 \left(1 -3 {(v\cdot q)^2\over q^2}+  6{(v\cdot q)^4\over q^4}\right)\right],
\end{eqnarray}
for the differential cross section in the black hole rest frame $v^\mu=(1,{\vec 0})$, as a function of the momentum transfer $q^\mu=p^\mu-p'^\mu$.   The point of this result is that, in generic regions of phase space, the integrated cross section scales as $\sim q^2/m^2_{Pl}$ relative to the leading order classical gravitational scattering between point masses $M$, $m_\phi$.   It is  parametrically of the same size as the sort of ${\cal O}(\hbar)$ corrections from graviton vacuum polarization loops that appear in Eq.~(\ref{eq:vhbar})~\cite{Goldberger:2020geb}.   Eq.~(\ref{eq:result}) can be therefore be regarded as a specific realization of a qualitatively new  phenomenon in low energy quantum gravity associated with the exchange of virtual Hawking gravitons and tractable (calculable) by the methods of EFT applied to gravity.

\section{Outlook}

Spurred by both theoretical and experimental breakthroughs, the first quarter of the 21st century has seen sustained progress in our understanding of the classical and quantum dynamics of gravity in the infrared.    On the theoretical front, this progress has been predicated in large part on the fact that in gravity, as in other field theories (\eg~ the Standard Model), UV physics decouples from long distance observables, so that the tools of effective field theory can be applied to gain conceptual insight and computational advantage.

As discussed in the introduction, the tools of EFT have been brought to bear in recent years on a number of questions in cosmology and astrophysics, where gravitational interactions play a crucial role.   An example that illustrates many of the new concepts that arise in the application of EFT ideas to gravity is the theory of compact objects first introduced in~\cite{Goldberger:2004jt}, which is the main focus of this review.   This tower of gravity EFTs gives a complete description of the adiabatic inspiral phase of compact binaries, systematically incorporating the effects of physics at scales ranging from the Schwarzschild radius up to the wavelength of the emitted gravitational waves.

This paper has focused on theories of gravity that flow to the Einstein-Hilbert Lagrangian in the infrared.    There exists a gigantic body of literature on long distance modifications of general relativity, either by relevant (e.g. mass) deformations  of Eq.~(\ref{eq:eh}) or by additional light fields in the gravitational sector.   A review of this literature is beyond the scope of this review.   However, many of the same EFT ideas mentioned here can be applied in such models.   In the case of binary dynamics, such EFTs have been analyzed e.g., in refs.~\cite{deRham:2005jan,Chu:2006ce,Porto:2007pw,deRham:2007mcp,Cardoso:2008gn,Hui:2009kc,Cannella:2009he,Gilmore:2009ea,Sanctuary:2010dao,deRham:2012fw,deRham:2012fg,Andrews:2013qva,Endlich:2017tqa,Baumann:2018vus,Dar:2018dra,Kuntz:2019zef,Brax:2019tcy,Wong:2019yoc,Kuntz:2019plo,Baumann:2019ztm,Sennett:2019bpc,Kuntz:2020gan,Kuntz:2020iie,Gonzalez:2020krh,Brax:2020vgg,Lins:2020omt,Brax:2021qqo}, while an EFT for extended defects of various dimensions coupled to gravity was introduced in~\cite{Emparan:2009cs,Emparan:2009at}.    

Another topic not treated in this review, which has been the subject of considerable recent efforts, is the interface between EFT methods for gravitational two-body dynamics and other perturbative approaches.   For instance, the  applications of EFT to the self-force problem in classical general relativity (see refs.~\cite{Poisson:2011nh,Barack:2018yvs} for reviews) have been discussed in~\cite{Galley:2008ih,Galley:2010xn,Galley:2011te,Galley:2013eba,Zimmerman:2015rga}, although this subject remains somewhat unexplored.   Finally, refs.~\cite{Neill:2013wsa,Vaidya:2014kza,Cheung:2018wkq} established a dictionary between the conservative sector of NRGR and the on-shell $S$-matrix for heavy particles coupled to gravity.   This has stimulated a flurry of theoretical activity, reviewed in ref.~\cite{Buonanno:2022pgc}.    Given the promising experimental situation in gravitational wave astronomy, it is a safe bet to expect sustained growth  in the years to come at the confluence of on-shell amplitudes, classical general relativity, and the EFT approach to compact binary mergers

\section{Acknowledgements}

The author thanks P. Draper and I. Rothstein for their patience, and R. Porto for comments on the manuscript.   This work was supported in part by the US Department of Energy under Grant No. DE­SC0017660.








\end{document}